\documentstyle[prb,aps,multicol,epsf]{revtex}

\begin{document}
\draft

\title{   Orbital dynamics in ferromagnetic transition metal oxides }

\author { Jeroen van den Brink, Peter Horsch, Frank Mack,
          and Andrzej M. Ole\'{s}\cite{AMO} }
\address{ Max-Planck-Institut f\"ur Festk\"orperforschung,
          Heisenbergstrasse 1, D-70569 Stuttgart,
          Federal Republic of Germany }

\date{September 22, 1998}

\maketitle

\begin{abstract}
We consider a model of strongly correlated $e_g$ electrons interacting by 
superexchange orbital interactions in the ferromagnetic phase of LaMnO$_3$. 
It is found that the classical orbital order with alternating occupied $e_g$ 
orbitals has a full rotational symmetry at orbital degeneracy, and the 
excitation spectrum derived using the linear spin-wave theory is gapless. 
The quantum (fluctuation) corrections to the order parameter and to the 
ground state energy restore the cubic symmetry of the model. By applying a 
uniaxial pressure orbital degeneracy is lifted in a tetragonal field and one 
finds an orbital-flop phase with a gap in the excitation spectrum. 
In two dimensions the classical order is more robust near the orbital 
degeneracy point and quantum effects are suppressed. The orbital excitation 
spectra obtained using finite temperature diagonalization of two-dimensional 
clusters consist of a quasiparticle accompanied by satellite structures. 
The orbital waves found within the linear spin-wave theory provide an 
excellent description of the dominant pole of these spectra. 
\end{abstract}


\begin{multicols}{2} 

\section{Introduction}
\label{sec:intro}

Recently much attention has been attracted to the properties of
La$_{1-x}$A$_x$MnO$_3$, with A=Ca,Sr,Ba, and related compounds, in
which the colossal magnetoresistance and metal-insulator transition are
observed.\cite{Ram97} The parent compound, LaMnO$_3$, is insulating with
layered-type antiferromagnetic (AF) ordered state (the A-type
antiferromagnet), accompanied by the orbital order with alternation of
$e_g$ orbitals occupied by a single electron of Mn$^{3+}$ ions.\cite{Goo55} 
Such orbitally ordered states might be promoted by a cooperative Jahn-Teller 
effect,\cite{Hal71} playing a decisive role in charge transport,\cite{Mil96} 
but could also result from an electronic instability.\cite{Var96} The latter 
possibility is now under debate, as the band structure calculations performed 
within the extension of local density approximation (LDA) to strongly 
correlated transition metal oxides, the so-called LDA+U approach, give
indeed charge-ordered ground states without assuming lattice distortions as 
their driving force.\cite{Sol96} Charge ordering was also found within the 
Hartree-Fock calculations on lattice models by Mizokawa and 
Fujimori.\cite{Miz96}

The dominating energy scale in late transition metal oxides is the local
Coulomb interaction $U$ between $d$ electrons at transition metal ions which 
motivates the description of ground state and low-energy excitations in terms 
of effective models with spin and orbital degrees of freedom. Such models, 
introduced both for the cuprates,\cite{Kug73,Fei97} and more recently for the
manganites,\cite{Ish96,Shi97,Fei98} show that the spin and orbital degrees of 
freedom are interrelated which leads to an interesting problem, even without 
doping. Orbital interactions in the ground state lead to a particular orbital 
ordering that depends on the magnetic ordering, and vice versa.
An interesting aspect of such system is that the elementary excitations
have either a pure spin, or pure orbital, or mixed spin-orbital 
character.\cite{Foz98,Bri98,Mil98} This gives large quantum fluctuation 
corrections to the order parameter in AF phases, and it was discussed recently 
that this might lead to a novel type of a spin-liquid.\cite{Fei97,Kha97,Foz98} 
In a special case of ferromagnetic (FM) order (which represents a simplified 
spinless problem), only orbital excitations contribute to the properties of 
the ground state and the quantum effects are expected to be smaller. 

Unlike in the cuprate materials, hole doping in the manganites leads to FM 
states, insulating at low doping, and metallic at higher doping.\cite{Sch95} 
Although the FM phase is unstable for the undoped LaMnO$_3$, the building 
blocks are FM planes. A proper understanding of pure orbital excitations has 
to start from the uniform FM phase, where the spin operators can be integrated 
out. Such a state could possibly be realized in LaMnO$_3$ at large magnetic 
field and will serve as a reference state to consider hole propagation in 
doped manganites, similarly as the AF spin state provides a reference state 
for the hole propagation in the cuprates. Therefore, we consider in this paper 
an effective model with only orbital superexchange interactions which involves 
the $e_g$ states and results from the effective Hamiltonians derived in 
Refs. \onlinecite{Fei97,Ish96,Shi97,Fei98} in the relevant high-spin state. 
Of course, it is identical for the cuprates and for the manganites. We note,
however, that orbital interactions may be also induced by lattice 
distortions, but these contributions are expected to be less important than
the orbital superexchange,\cite{Lic95} and will be neglected.

So far, little is known about the consequences of orbital excitations for 
long-range order and transport properties in doped systems, but it might be 
expected that such excitations either bind to a moving hole, or lead to hole 
scattering even when the spins are aligned.\cite{Zaa93} The knowledge of 
orbital excitations is a prerequisite for a better understanding of the 
temperature dependence of the resistivity and optical conductivity in the FM 
phase of doped manganites. Although the double exchange mechanism is quite 
successful in explaining why the spin order becomes FM with increasing doping 
of La$_{1-x}$A$_x$MnO$_3$, it fails to reproduce the experimentally observed 
temperature dependence of the resistivity.\cite{Mil95} Even more puzzling and 
contradicting naive expectations is the incoherent optical conductivity with 
a small Drude peak observed in a FM metallic phase at higher 
doping.\cite{Oki97} This behavior demonstrates the importance of orbital 
dynamics in doped manganites which appears to be predominantly incoherent in 
the relevant parameter regime.\cite{Hor98,Kil98} 

The paper is organized as follows. The orbital superexchange model is
presented in Sec. II. We derive the classical phases with alternating
orbitals on two sublattices in three and two dimensions, and analyze their 
dependence on the crystal field splitting. Orbital excitations are derived 
in Sec. III using a pseudospin Hamiltonian and an extension of the linear 
spin-wave theory (LSWT) to the present situation with no conservation of 
orbital quantum number. We present numerical results for the dispersion 
of orbital waves and determine quantum fluctuation corrections to the 
ground-state energy and to the order parameter in Sec. IV. 
The derived spectra are compared with the results of exact diagonalization 
in Sec. V. A short summary and conclusions are given in Sec. VI.

\section{Orbital Hamiltonian and classical states}
\label{sec:model}

\subsection{ Orbital superexchange interactions }
\label{sec:exchange}

\begin{figure}
      \epsfysize=50mm
      \centerline{\epsffile{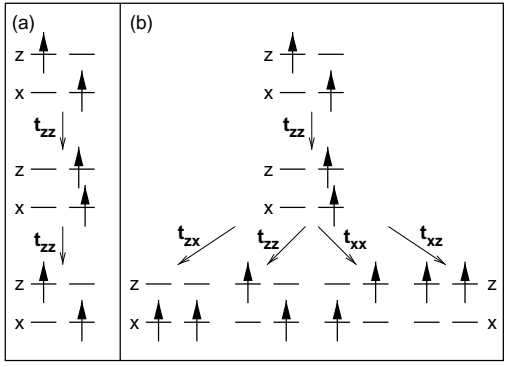}}
\caption
{Schematic representation of the virtual $d_i^4d_j^4\rightarrow d_i^3d_j^5$ 
 excitations in LaMnO$_3$ for the starting FM configuration
 $d_{iz\uparrow}^{\dagger}d_{jx\uparrow}^{\dagger}|0\rangle$ which involve 
 the high-spin $|^6\!A_1\rangle$ state and generate effective orbital
 superexchange interactions:
 (a) for a bond along the $c$-axis, $(ij)\perp$;
 (b) for a bond within the $(a,b)$ plane, $(ij)\parallel$. }
\label{fig:scheme}
\end{figure}

We consider a three-dimensional (3D) Mott insulator with one $e_g$ electron 
per site. The effective Hamiltonian which describes $e_g$ electrons in 
a cubic crystal at strong-coupling,\cite{Fei97,Ish96,Shi97,Fei98}
\begin{equation}
{\cal H} = H_J + H_z,
\label{somfff}
\end{equation}
consists of the superexchange part $H_J$, and the orbital splitting term due 
to crystal-field term $H_z$. Here we will consider a special case of the FM 
$e_g$ band in which the spin dynamics is integrated out and one is left with 
the effective interactions between the electrons (holes) in different orbital 
states. In this case the only superexchange channel which contributes is the 
effective interaction via the high-spin state, and thus the interaction 
occurs only between the pairs of ions with singly occupied orthogonal $e_g$ 
orbitals at two nearest-neighbor sites (two orthogonal $e_g$ orbitals are 
then singly occupied in the intermediate excited states). An example of such 
a model is the superexchange interaction in the FM state of LaMnO$_3$ which 
originates from $d_i^4d_j^4\rightleftharpoons d_i^3d_j^5$ excitations (with
$d_i^4\equiv t_{2g}^3e_g$, $d_i^3\equiv t_{2g}^3$, and 
$d_i^4\equiv t_{2g}^3e_g^2$) into a high-spin $d_j^5$ state $|^6\!A_1\rangle$, 
as shown schematically in Fig. \ref{fig:scheme}. It gives the effective 
Hamiltonian with orbital interactions,\cite{Shi97,Fei98}
\begin{equation}
H_J = - \frac{t^2}{\varepsilon(^6\!A_1)}
        \sum_{\langle ij\rangle} {\cal P}_{\langle ij\rangle}^{\zeta\xi},
\label{generic}
\end{equation}
where $t$ is the hopping element between the directional $3z^2-r^2$
orbitals along the considered $c$-axis, and $\varepsilon(^6\!A_1)$ is
the excitation energy. The orbital degrees of freedom are described by
the projection operators ${\cal P}_{\langle ij\rangle}^{\zeta\xi}$ which
select a pair of orbitals $|\zeta\rangle$ and $|\xi\rangle$, being
parallel and orthogonal to the directions of the considered bond
$\langle ij\rangle$ in a cubic lattice. 

The Hamiltonian (\ref{generic}) has cubic symmetry and may be written 
using any reference basis in the $e_g$ subspace. For the conventional choice 
of $3z^2-r^2\sim |z\rangle$ and $x^2-y^2\sim |x\rangle$ orbitals, the above
projection operators are represented by the orbital operators
$\tau_i^{\alpha}$, with $\alpha=a,b,c$ for three cubic axes,
\begin{equation}
{\cal P}_{\langle ij\rangle}^{\zeta\xi} =
  (\case{1}{2}-\tau_i^{\alpha}) (\case{1}{2}+\tau_j^{\alpha})
+ (\case{1}{2}+\tau_i^{\alpha}) (\case{1}{2}-\tau_j^{\alpha}).
\label{projection}
\end{equation}
It is convenient to replace the orbital operators $\tau_i^{\alpha}$ by the
{\em pseudospin\/} operators $T^{\mu}_i$ with $\mu=x,z$,
\begin{equation}
\tau^{a(b)}_i = -\case{1}{2}( T^z_i \mp\sqrt{3}T^x_i ), \hskip .7cm
\tau^c_i      =  T^z_i,
\label{eq:orbop}
\end{equation}
The latter operators may be represented by the Pauli matrices in the same 
way as spin operators, $T^{\alpha}_i=\frac{1}{2}\sigma^{\alpha}_i$, and obey 
the same commutation relations. Therefore, the $z$th component of pseudospin 
is given by $T^z_i=\frac{1}{2}(n_{ix}-n_{iz})$, and we identify the orbital
states as up- and down-pseudospin, $|x\rangle\equiv |\uparrow\rangle$, and
$|z\rangle\equiv |\downarrow\rangle$, respectively. We take the prefactor 
$J=t^2/\varepsilon(^6\!A_1)$ in Eq. (\ref{generic}) as the energy unit for 
the superexchange interaction. Thus, one finds a pseudospin Hamiltonian,
\begin{eqnarray}
H_J &=& \case{1}{2}J \sum_{\langle ij\rangle\parallel} \left[
        T^z_iT^z_j+3T^x_iT^x_i\mp\sqrt{3}( T^x_iT^z_j+T^z_iT^x_j )\right]
           \nonumber \\
    &+&  2J \sum_{\langle ij\rangle\perp}  T^z_i T^z_j ,
\label{hamorb}
\end{eqnarray}
where the prefactor of the mixed term $\propto\sqrt{3}$ is negative in
the $a$-direction and positive in the $b$-direction.\cite{notephase} We
choose a convention that the bonds labeled as
$\langle ij\rangle\!\!\parallel$ ($\langle ij\rangle\!\!\perp$) connect
nearest-neighbor sites within $(a,b)$ planes (along the $c$-axis). The
virtual excitations which lead to the interactions described by Eq.
(\ref{hamorb}) are shown in Fig. \ref{fig:scheme}. Here we neglected a trivial
constant term which gives the energy of $-J/2$ per bond, i.e., $-3J/2$
per site in a 3D system. We emphasize that the $SU(2)$ symmetry is explicitly 
broken in $H_J$, and the interaction depends only on two pseudospin operators, 
$T^x_i$ and $T^z_i$.

The crystal-field term removes the degeneracy of $|x\rangle$ and
$|z\rangle$ orbitals,
\begin{equation}
H_z = - E_z \sum_i T_i^z,
\label{hz}
\end{equation}
and is connected with the uniaxial pressure. This term is induced by static 
distortions in a tetragonal field and is of particular importance in 
a two-dimensional (2D) system.

\subsection{ Classical ground states }
\label{sec:3D}

Before analyzing the excitation spectra of the Hamiltonian (\ref{hamorb}), 
one has to determine 
first the classical ground state of the system that will serve as a reference 
state to calculate the Gaussian fluctuations. Neglecting the irrelevant phase
factors, the classical configurations which minimize the interaction terms 
(\ref{hamorb}) are characterized by the two-sublattice pseudospin order, with 
two angles describing orientations of pseudospins, one at each sublattice. 
As usually, the classical ground state is obtained by minimizing the energy 
with respect to these two rotation angles, i.e., by choosing the optimal 
orbitals.

Let us consider first the term $H_J$ at orbital degeneracy $E_z=0$. The 
superexchange interaction (\ref{generic}) induces the alternation of 
{\em orthogonal orbitals\/} in the ground state in all three directions, 
and is equivalent to the {\it two-sublattice AF ({\rm G-AF}) order in the 
pseudospin space\/}. This configuration gives the lowest energy on the 
mean-field level, as the virtual transitions represented in Fig. 1 give the 
largest contribution, if the hopping involves one occupied and one unoccupied 
orbital of the same type (e.g., either directional or planar). 
{\it A priori\/}, the optimal choice of the occupied orbitals could be unique 
on the classical level due to the cubic symmetry of superexchange interaction 
(\ref{generic}). Therefore, we perform a uniform rotation of 
$\{|z\rangle,|x\rangle\}$ orbitals by an angle $\theta$ at each site, 
\begin{equation}
\left( \begin{array}{c}
 |i\bar{\mu}\rangle   \\
 |i\bar{\nu}\rangle
\end{array} \right) =
\left(\begin{array}{cc}
 \ \ \cos\theta &  \sin\theta \\
-\sin\theta &  \cos\theta
\end{array} \right)
\left( \begin{array}{c}
 |iz\rangle   \\
 |ix\rangle
\end{array} \right) ,
\label{newstatei}
\end{equation}
to generate the new orthogonal orbitals, $|i\bar{\mu}\rangle$ and 
$|i\bar{\nu}\rangle$, and investigate the energy as a function of $\theta$. 
The AF order imposed by the Hamiltonian $H_J$ in a cubic system implies the 
alternation of these orthogonal orbitals in the rotated basis, 
$|i\bar{\mu}\rangle$ and $|j\bar{\nu}\rangle$, at two sublattices, i.e., 
$i\in A$ and $j\in B$. A few examples of various possible choices of the 
orthogonal orbitals $\{|i\bar{\mu}\rangle,|i\bar{\nu}\rangle\}$ are given 
in Table I for representative values of $\theta$.

The rotation (\ref{newstatei}) leads to the following transformation of the 
pseudospin operators, 
\begin{eqnarray}
T^x_i &\rightarrow&  T^x_i \cos 2\theta - T^z_i \sin 2\theta,  \nonumber\\
T^z_i &\rightarrow&  T^x_i \sin 2\theta + T^z_i \cos 2\theta,
\label{eq:rot}
\end{eqnarray}
and the interaction Hamiltonian $H_J$ is then transformed into,
\begin{equation}
{\cal H}^\theta = H_\parallel^\theta + H_\perp^\theta, 
\label{hrot}
\end{equation}
\begin{eqnarray}
H_\parallel^\theta &=&  \frac{J}{2} \sum_{\langle ij\rangle\parallel} 
  \left[\right. (2+\cos 4\theta\mp \sqrt{3}\sin 4\theta ) T^x_i T^x_j 
                                                          \nonumber \\
 & &\hskip 1.1cm +(2-\cos 4\theta\pm\sqrt{3}\sin 4\theta )T^z_i T^z_j  
                                                          \nonumber \\
 & & \hskip 0.6cm -(\sin 4\theta\pm\sqrt{3}\cos 4\theta ) 
                      (T^x_iT^z_j + T^z_iT^x_j) \left. \right],    
\label{hrotpara}
\end{eqnarray}
\begin{eqnarray}
H_\perp^\theta &=& J\sum_{\langle ij\rangle\perp} \left[ \right.
      (1-\cos 4\theta) \ T^x_i T^x_j + (1+\cos 4\theta) \ T^z_i T^z_j 
                                                          \nonumber \\
    & & \hskip 1.1cm + \sin 4\theta (T^x_iT^z_j + T^x_iT^z_j)\left. \right]. 
\label{hrotperp}
\end{eqnarray}
As in Eq. (\ref{hamorb}), the bonds in the first sum (\ref{hrotpara}) are 
parallel to either $a$ or $b$-axis, while the bonds in the second sum 
(\ref{hrotperp}) are parallel to the $c$-axis. The classical energy is 
minimized if $\langle T^z_i T^z_j\rangle=-\frac{1}{4}$, i.e., if the 
orbitals order 'antiferromagnetically'. 

The Hamiltonian given now by Eqs. (\ref{hrotpara}) and (\ref{hrotperp})
has the symmetry of the cubic lattice, but {\em surprisingly one finds 
the full rotational symmetry\/} of the present interacting problem on the 
classical level at orbital degeneracy. In other words, classically the 
lowest energy is always $E_{\rm MF}=-3J/4$ per site, independent of the 
rotation angle, as long as alternating orbitals on neighboring sites are 
occupied. This follows from the particular structure of the rotated 
Hamiltonian (\ref{hrot}) which has identical factors of three in front of 
the diagonal $\propto T^z_i T^z_j$ and off-diagonal $\propto T^x_i T^x_j$ 
contributions, when these are summed over all the bonds which originate at 
each site, and these coefficients are independent of the rotation angle 
$\theta$. In contrast to the Heisenberg antiferromagnet (HAF), however, 
this symmetry concerns only the {\em summed\/} contributions, and not the 
interactions along the individual bonds.   

\begin{figure}
      \epsfxsize=80mm
      \centerline{\epsffile{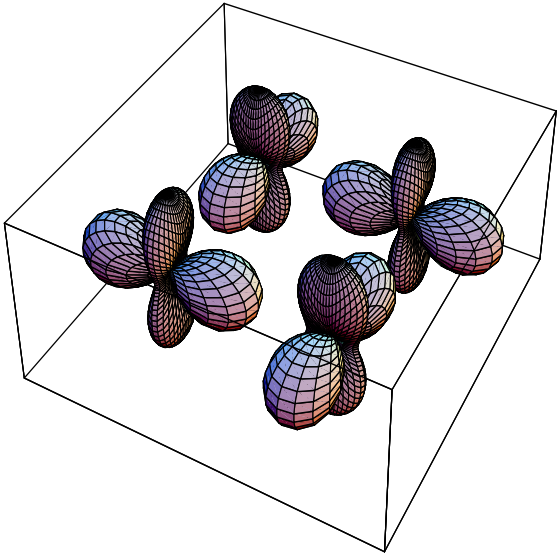}}
\caption{
 Alternating orbital order in FM LaMnO$_3$: 
 $(|x\rangle + |z\rangle )/\sqrt{2}$ an
 $(|x\rangle - |z\rangle )/\sqrt{2}$ as found at $E_z\to 0$. }
\label{fig:orbitals}
\end{figure}

A finite orbital field $E_z\neq 0$ breaks the rotational symmetry on the
classical level. It acts along the $c$-axis, and it is therefore easy to 
show that the ground state in the limit of $E_z\to 0$ is realized by the 
alternating occupied orbitals being symmetric/antisymmetric linear 
combinations of $|z\rangle$ and $|x\rangle$ orbitals,\cite{Fei98} i.e., 
the occupied states correspond to the rotated orbitals (\ref{newstatei}) 
$|i\bar{\mu}\rangle$ and $|i\bar{\nu}\rangle$ on the two sublattices with an 
angle $\theta=\pi/4$, shown in Fig. \ref{fig:orbitals} (see also Table I). 
In particular, this state 
is different from the alternating directional orbitals, $3x^2-r^2$ and 
$3y^2-r^2$, which might have been naively expected. It follows in the 
limit of degenerate orbitals from the 'orbital-flop' phase, in analogy to 
a spin-flop phase for the HAF at finite magnetic field. With increasing 
(decreasing) $E_z$ the orbitals tilt out of the state shown in Fig.
\ref{fig:orbitals}, and approach $|x\rangle$ $(|z\rangle)$ orbitals, 
respectively, which may be interpreted as an increasing FM component of the 
magnetization in the pseudospin model. 

It is convenient to describe this tilting of pseudospins due to the
crystal-field $\propto E_z$ by making two different transformations 
(\ref{newstatei}) at both sublattices, rotating the orbitals by an angle 
$\theta=\case{\pi}{4}-\phi$ on sublattice $A$, and by an angle 
$\theta=\case{\pi}{4}+\phi$ on sublattice $B$, so that the relative angle 
between the {\em occupied\/} orbitals is $\case{\pi}{2}-2\phi$ 
and decreases with increasing $\phi$, i.e., with increasing $E_z$,
\begin{equation}
\left( \begin{array}{c}
 |i\mu\rangle   \\
 |i\nu\rangle
\end{array} \right) \! = \!
\left(\begin{array}{cc}
 \ \ \cos(\case{\pi}{4}-\phi) & \sin(\case{\pi}{4}-\phi) \\
    -\sin(\case{\pi}{4}-\phi) & \cos(\case{\pi}{4}-\phi)
\end{array} \right)\!
\left( \begin{array}{c}
 |iz\rangle   \\
 |ix\rangle
\end{array} \right) ,
\label{flopi}
\end{equation}
\begin{equation}
\left( \begin{array}{c}
 |j\mu\rangle   \\
 |j\nu\rangle
\end{array} \right) \! = \!
\left(\begin{array}{cc}
 \ \ \cos(\case{\pi}{4}+\phi) & \sin(\case{\pi}{4}+\phi) \\
    -\sin(\case{\pi}{4}+\phi) & \cos(\case{\pi}{4}+\phi)
\end{array} \right)\!
\left( \begin{array}{c}
 |jz\rangle   \\
 |jx\rangle
\end{array} \right) .
\label{flopj}
\end{equation}
As before, the orbitals $|i\mu\rangle$ and $|j\nu\rangle$ are occupied at 
two sublattices, $i\in A$ and $j\in B$, respectively, and the orbital order 
is AF in the classical state, with the transformed operators 
$\langle T^z_i\rangle=-1/2$ and $\langle T^z_j\rangle=+1/2$, respectively. 
As a result one finds that the angles for the occupied orbitals are indeed 
opposite on the two sublattices. 

The operators $T^x_i$ and $T^z_i$ may be now transformed as in Eqs. 
(\ref{eq:rot}) using the actual rotations by $\theta=\case{\pi}{4}\pm\phi$,
as given in Eqs. (\ref{flopi}) and (\ref{flopj}), and
the transformed Hamiltonian takes the form,
\begin{equation}
{\cal H}^\phi=H_\parallel^\phi+H_\perp^\phi+H_z^\phi,\\
\label{hflop}
\end{equation}
\begin{eqnarray}
H_\parallel^\phi &=& \frac{J}{2}\sum_{\langle ij\rangle\parallel}\left[ 
         (2\cos 4\phi -1) T^x_i T^x_j 
        +(2\cos 4\phi +1) T^z_i T^z_j \right. \nonumber \\
    & & \hskip 0.9cm +2\sin 4\phi(T^x_i T^z_j-T^z_i T^x_j) \nonumber \\
    & & \hskip 0.9cm\pm\! \sqrt{3} (T^x_i T^z_j+T^z_i T^x_j)\left.\right], 
\label{hfloppara}
\end{eqnarray}
\begin{eqnarray}
H_\perp^\phi &=& J\sum_{\langle ij\rangle\perp}\left[ \right.
            (\cos 4\phi +1)T^x_iT^x_j +(\cos 4\phi -1)T^z_iT^z_j  \nonumber \\
    & & \hskip 0.9cm -\sin 4\phi( T^x_i T^z_j - T^z_i T^x_j )\left.\right],  
\label{hflopperp}
\end{eqnarray}
\begin{equation}
H_z^\phi = E_z\sum_i (\lambda_i\sin 2\phi T^z_i - \cos 2\phi T^x_i),
\label{hflopez}
\end{equation}
where $\lambda_i=-1$ for $i\in A$ and $\lambda_i=1$ for $i\in B$. The energy  
of the classical ground state is given by,
\begin{equation}
E_{\rm 3D}^{\rm MF}=-\frac{3}{4}J\cos 4\phi
                    -\frac{1}{2}E_z\sin 2\phi,
\label{emf3d}
\end{equation}
and is minimized by,
\begin{equation}
\sin 2\phi=\frac{E_z}{6J}.
\label{cos3d}
\end{equation}
At orbital degeneracy ($E_z=0$) this result is equivalent to $\cos 2\theta=0$
($\theta=\pi/4$) in Eq. (\ref{newstatei}), as obtained also in Ref. 
\onlinecite{Fei98}, and one recovers from Eq. (\ref{emf3d}) the energy of 
$-3J/4$ as a particular realization of the degenerate classical phases with 
alternating orthogonal orbitals. The above result (\ref{cos3d}) is valid for 
$|E_z|\leq 6J$; otherwise one of the initial orbitals (either $|x\rangle$ or 
$|z\rangle$) is occupied at each site, and the state is fully polarized 
($\sin 2\phi=\pm 1$). The case of $\phi=0$ corresponds to the occupied 
orthogonal orbitals at orbital degeneracy. In contrast, the value of 
$\phi=\pi/4$ corresponds to performing no rotation on the $A$ sublattice, 
while the orbitals are interchanged on the $B$ sublattice [see Eqs. 
(\ref{flopi}) and (\ref{flopj})]. The latter situation describes a state 
obtained in strong orbital field, with only one type of orbitals occupied, 
a 'FM orbital state'. The orbital field changes therefore the orbital 
ordering from the AF to FM one in a continuous fashion. 

\subsection{Two-dimensional orbital model}
\label{sec:2D}

As a special case we consider also a 2D orbital model with the interactions 
in the $(a,b)$ plane. In this case the cubic symmetry is explicitly broken, 
and the classical state is of a spin-flop type. It corresponds to 
alternatingly occupied orbitals on the two sublattices, with the orbitals
oriented in the plane and given by $\theta=\pi/4$ at $E_z=0$, see Fig. 
\ref{fig:orbitals}.\cite{noteexp} A finite value of $E_z$ tilts the orbitals 
out of the planar $|x\rangle$ orbitals by an angle $\phi$, and the Hamiltonian 
reduces to
\begin{eqnarray}
{\cal H}^\phi_{2D}=H_\parallel^\phi+H_z^\phi,
\end{eqnarray}
as there is no bond in the $c$-direction. The classical energy is
\begin{equation}
E_{\rm 2D}^{\rm MF}=-\frac{1}{4}J(2\cos 4\phi+1)
                    -\frac{1}{2}E_z\sin 2\phi.
\label{emf2d}
\end{equation}
Therefore, one finds the same energy of $-3J/4$ as in a 3D model at
orbital degeneracy. This shows that the {\em orbital superexchange
interactions are geometrically frustrated\/}, and the bonds in the third
dimension cannot lower the energy, but only allow for restoring the
rotational symmetry on the classical level and rotating the orthogonal
orbitals in an arbitrary way. The energy (\ref{emf2d}) is minimized by,
\begin{equation}
\sin 2\phi=\frac{E_z}{4J},
\label{cos2d}
\end{equation}
if $|E_z|\leq 4J$; otherwise $\sin 2\phi=\pm 1$.
Interestingly, the value of the field at which the orbitals are fully
polarized is reduced by one third from the value obtained in three dimensions 
(\ref{cos3d}). This shows that although the orbital exchange energy can 
be gained in a 3D model only on the bonds along two directions in the
alternating (orbital-flop) phase either at or close to $E_z=0$, one has to 
counteract the superexchange on the bonds in all three directions when the
field is applied.

We note that the present orbital model becomes purely classical in one
dimension, if the lattice effects are neglected, as we have assumed in the 
present model. This follows from the derivation which gives the 
superexchange interactions of Ising type between the $|z\rangle$ orbitals 
along the $c$-axis, and one can always choose this axis along the considered 
chain. Thus, one finds the Hamiltonian
\begin{equation}
{\cal H}_{\rm 1D}=2J\sum_i T_i^zT_{i+1}^z-E_z\sum_i T_i^z.
\label{h1d}
\end{equation}
At orbital degeneracy ($E_z=0$) the occupied orbitals alternate along 
the one-dimensional (1D) chain between two orthogonal orbitals, and one
may for example choose the occupied states as $|z\rangle$ and $|x\rangle$ 
on even and odd sites, respectively. The energy of this ground state per 
site is $-J/2$, which allows to conclude that energy gains due to orbital 
ordering per site are significantly reduced compared to the 2D and 3D case.

\section{Orbital excitations in linear spin-wave theory}
\label{sec:lsw}

The superexchange in the orbital subspace is AF and one may map the orbital 
terms in the Hamiltonian (\ref{hamorb}) onto a spin problem in order to treat 
the elementary excitations within the LSWT. The equations of motion can then
be linearized by a standard technique.\cite{Hal71} Let us first discuss the 
results of LSWT for the spin-flop phase induced by an orbital-field, starting
from the rotated Hamiltonian (\ref{hflop}). Here we choose the 
Holstein-Primakoff transformation \cite{Aue94} for localized pseudospin 
operators ($T=1/2$),
\begin{eqnarray}
T_i^+ &=& \bar{a}^{\dag}_i(1-\bar{a}_i^{\dag}\bar{a}_i)^{1/2},           \ \ 
T_i^-  =  (1-\bar{a}_i^{\dag}\bar{a}_i)^{1/2}\bar{a}_i,         \nonumber \\
T_i^z &=& \bar{a}^{\dag}_i\bar{a}_i-\frac{1}{2},
\label{hpi}
\end{eqnarray}
for $i\in A$ sublattice and
\begin{eqnarray}
T_j^+ &=& (1-\bar{b}_j^{\dag}\bar{b}_j)^{1/2}\bar{b}_j,                 \ \ 
T_j^-  =  \bar{b}^{\dag}_j(1-\bar{b}_j^{\dag}\bar{b}_j)^{1/2}, \nonumber \\
T_j^z &=& \frac{1}{2}-\bar{b}^{\dag}_j\bar{b}_j,
\label{hpj}
\end{eqnarray}
for $j\in B$ sublattice. After replacing the square roots by the leading 
lowest order terms, and inserting the expansion into Eq. (\ref{hflop}), one 
diagonalizes the linearized Hamiltonian by two consecutive transformations. 
Note that, contrary to the HAF, also cubic terms in boson operators are 
present in the expansion due to the anomalous interactions $\sim T_i^zT_j^x$. 

The linearized Hamiltonian simplifies by introducing the Fourier transformed 
boson operators $\bar{a}_{\bf k}$ and $\bar{b}_{\bf k}$ given by,
\begin{eqnarray}
\bar{a}_{\bf k}=\sqrt{2\over{N}}\sum_{i\in A}e^{i{\bf k}{\bf r}_i}\bar{a}_{i},
\ \ \  
\bar{b}_{\bf k}=\sqrt{2\over{N}}\sum_{j\in B}e^{i{\bf k}{\bf r}_j}\bar{b}_{j}.
\nonumber
\end{eqnarray}
and by transforming to new boson operators $\{a_{\bf k},b_{\bf k}\}$,
\begin{eqnarray}
a_{\bf k} = (\bar{a}_{\bf k} - \bar{b}_{\bf k})/\sqrt{2}  ,\ \ \ 
b_{\bf k} = (\bar{a}_{\bf k} + \bar{b}_{\bf k})/\sqrt{2}.  
\label{eq:branch}
\end{eqnarray}
One finds the effective orbital Hamiltonian of the form,
\begin{eqnarray}
H_{\rm LSW} &=& J \sum_{\bf k} \left[ A_{\bf k}a^{\dag}_{\bf k}a_{\bf k}
+ \frac{1}{2} B_{\bf k}(a^{\dag}_{\bf k}a^{\dag}_{-\bf k}
+ a^{}_{\bf k}a^{}_{-\bf k}) \right] \nonumber \\
&+& J \sum_{\bf k} \left[ A_{\bf k}b^{\dag}_{\bf k}b_{\bf k}
- \frac{1}{2} B_{\bf k}(b^{\dag}_{\bf k}b^{\dag}_{-\bf k}
+ b^{}_{\bf k}b^{}_{-\bf k}) \right] ,
\label{hfou}
\end{eqnarray}
where the coefficients $A_{\bf k}$ and $B_{\bf k}$ depend on angle $\phi$,
\begin{eqnarray}
A_{\bf k}&=&3-B_{\bf k},    \\
B_{\bf k}&=&\frac{1}{2}\left[(2\cos 4\phi-1)\gamma_{\parallel}({\bf k})
          + (\cos 4 \phi+1)\gamma_{\perp}({\bf k}) \right] ,         
\label{eq:gamma}  
\end{eqnarray}
and the ${\bf k}$-dependence is given by 
$\gamma_{\parallel}({\bf k})=\frac{1}{2}(\cos k_x+\cos k_y)$, and by
$\gamma_{\perp}({\bf k})=\cos k_z$, respectively. After a Bogoliubov 
transformation,
\begin{equation}
\label{Bog}
a_{\bf k}=u_{\bf k}\alpha_{\bf k}+v_{\bf k}\alpha^{\dag}_{\bf -k},
\label{bt}
\end{equation}
with the parameters
\begin{eqnarray}
u_{\bf k}=\sqrt{\frac{A_{\bf k}}{2\eta_{\bf k}}+\frac{1}{2}},\ \ \ 
v_{\bf k}=-{\rm sgn}(B_{\bf k}) 
\sqrt{\frac{A_{\bf k}}{2\eta_{\bf k}}-\frac{1}{2}}, 
\label{uvlsw}
\end{eqnarray}
with $\eta_{\bf k}=\sqrt{A_{\bf k}^2-B_{\bf k}^2}$, and an equivalent 
transformation for the $b_{\bf k}$ bosons, the Hamiltonian (\ref{hfou}) is 
diagonalized, and takes the following form,
\begin{equation}
H_{\rm LSW}=\sum_{\bf k}\left[ 
\omega^-_{\bf k}(\phi)\alpha^{\dag}_{\bf k}\alpha_{\bf k} +
\omega^+_{\bf k}(\phi) \beta^{\dag}_{\bf k} \beta_{\bf k}\right] .  
\label{Hnu}
\end{equation}
The orbital-wave dispersion is given by  
\begin{eqnarray}
\omega_{\bf k}^{\pm}(\phi)&=&3J\left\{ 1 \pm \case{1}{3}\left[
 (2\cos4\phi-1)\gamma_{\parallel}({\bf k})\right.\right. \nonumber \\
& &\left.\left.\hskip .8cm + (\cos4\phi+1)\gamma_{\perp}({\bf k}) 
\right]\right\}^{1/2}.
\label{eq:omega_phi}
\end{eqnarray}
The orbital excitation spectrum consists of two branches like, for instance, 
in an anisotropic Heisenberg model.\cite{Kan55}

The dependence on the field $E_z$ is implicitly contained in the above 
relations via the angle $\phi$, as determined by Eqs. (\ref{cos3d}) and 
(\ref{cos2d}) for the 3D and 2D model, respectively. The dispersion changes 
between the case of $\phi=0$ ($\cos 4\phi=1$) which stands for the spin-flop 
phase at orbital degeneracy, and the case of $\phi=\pi/4$ ($\cos 4\phi=-1$) 
which corresponds to the uniform phase with either $|x\rangle$ ($E_z\geq 6J$) 
or $|z\rangle$ ($E_z\leq -6J$) orbitals occupied. The orbital field therefore 
changes the orbital-wave dispersion from AF to FM state.
The orbital-wave dispersion for a 2D system, can easily be found from
Eq. (\ref{eq:omega_phi}) by setting $\gamma_{\perp}({\bf k})=0$.

In the case of a vanishing orbital field the classical ground state is 
degenerate with respect to a rotation of the orbital around an arbitrary 
angle $\theta$. In LSWT, nevertheless, the orbital-wave dispersion depends on 
$\theta$, i.e., the orbital-wave velocity is highly anisotropic, and one finds,
\begin{eqnarray}
\Omega_{\bf k}^{\pm}(\theta) &=& 3J  \left\{  1 \pm \case{1}{6}\left[ 
   (\sqrt{3}\cos 2\theta -\sin 2\theta)^2\gamma_x({\bf k})\right.\right.
                                                                 \nonumber \\
&+&\left.\left.\!   
   (\sqrt{3}\cos 2\theta +\sin 2\theta)^2\gamma_y({\bf k})
+ 4\sin^2 2\theta\gamma_{\perp}({\bf k})
\right]\!\right\}^{\frac{1}{2}}\!\! ,    \nonumber \\
\label{eq:omega_theta}
\end{eqnarray}
with $\gamma_{x,y}({\bf k})=\cos k_{x,y}$ respectively. It is straightforward 
to verify that the excitations are gapless, independent of the rotation angle.
If $\theta=0$, the orbital-waves given by Eq. (\ref{eq:omega_phi}) take a 
particularly simple form, and is identical to the dispersion found in the 
spin-flop phase at $\phi=\pi/4$,
\begin{equation}
\Omega_{\bf k}^{\pm}(\theta=0)=3J\sqrt{1\pm\gamma_{\parallel}(\bf k)}
                              =\omega_{\bf k}^{\mp}(\phi=\pi/4).
\label{omegaxz}
\end{equation}
In both cases the system is quasi-2D and the dispersion originates only from 
the planar components $k_x$ and $k_y$.

\section{Numerical results}
\label{sec:numerical}

\begin{figure} 
      \epsfxsize=85mm
      \centerline{\epsffile{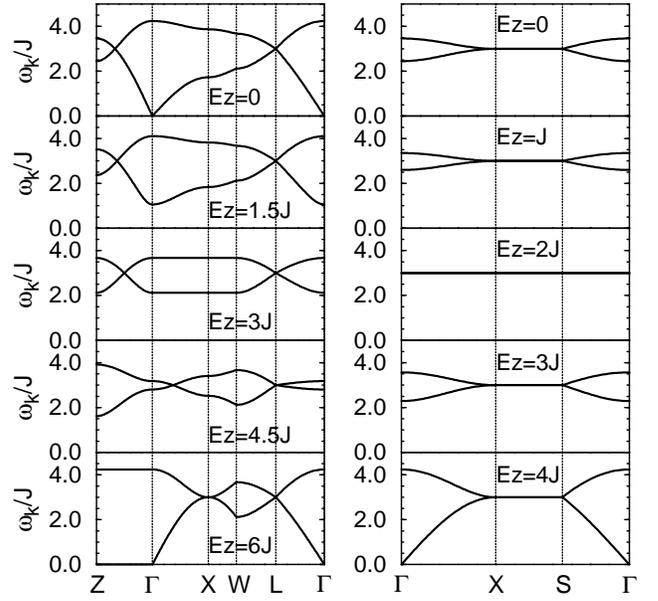}}
\caption
{Orbital-wave excitations as obtained for different values of
 the crystal-field splitting $E_z$ for a 3D (left) and 2D (right) orbital
 superexchange model (\protect{\ref{somfff}}). The result shown for a 3D
 system at $E_z=0$ was obtained for the orbitals rotated by $\theta=\pi/4$
 (\protect{\ref{newstatei}}),
 and corresponds to the $E_z\to 0$ limit of the orbital-flop phase. }
\label{fig:modes}
\end{figure}

\begin{figure}
      \epsfxsize=75mm
      \centerline{\epsffile{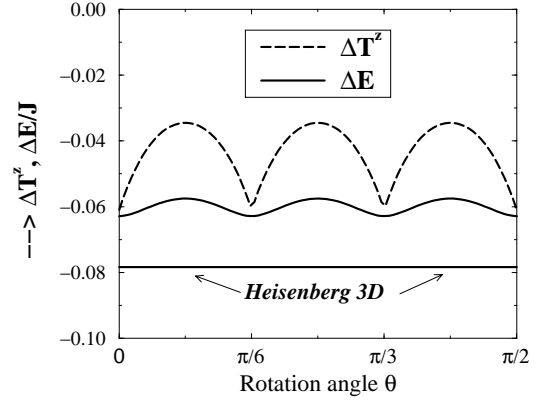}}
\vskip .5cm
\caption
{Quantum corrections for the 3D system as functions of rotation angle
 $\theta$ for: the renormalized order parameter $\Delta T^z$ (full lines), 
 and the ground-state energy $\Delta E/J$ (dashed lines). }
\label{fig:theta}
\end{figure}

The excitation spectra given by Eq. (\ref{eq:omega_phi}) for different
crystal-field splittings are shown in Fig. \ref{fig:modes} for the 3D  
system along different directions of the fcc Brillouin zone \cite{Kittel}
appropriate for the alternating orbital order.
Most interestingly, a gapless orbital-wave excitation is found for 
the 3D system at orbital degeneracy. Obviously this is due to the fact that 
the classical ground state energy is independent of the rotation angle 
$\theta$ at $E_z=0$, as we have shown in Sec. \ref{sec:model}. At first 
glance, however, one does not expect such a gapless mode, as the Hamiltonian 
(\ref{hamorb}) does not obey a continuous $SU(2)$ symmetry. The cubic 
symmetry of the model, however, is restored if one includes the quantum 
fluctuations; they are shown in Fig. \ref{fig:theta} as functions of the 
rotation angle $\theta$. Note that the quantum corrections are small and 
comparable to those of the 3D HAF. They do depend on the rotation angle, as 
the orbital-wave dispersion does. If the dispersion is completely 2D-like,
$\omega_{\bf k}^{\pm}(\phi=\pi/4)=3J\sqrt{1\pm\gamma_{\parallel}(\bf k)}$, 
corresponding to $\Omega_{\bf k}^{\pm}(\theta=0)$ (\ref{omegaxz}), the 
quantum corrections are largest, as this dispersion has a line of nodes along 
the $\Gamma-Z$ direction, i.e., $\omega^-_{(0,0,q)}=0$ for $0<q<\pi$. 

It is interesting to note that this fully 2D orbital-wave dispersion leads to 
an orbital ordered state at zero temperature only. At any finite temperature 
fluctuations destroy the long-range order (as in the 2D HAF). In this way 
one finds an orbital-liquid state at non-zero temperature, similar to that 
obtained in a Schwinger-boson approach by Ishihara, Yamanaka and Nagaosa for 
the doped system.\cite{Nag97} In higher order spin-wave theory, however, it 
might very well be that a gap occurs in the excitation spectrum. We expect, 
however, that this gap, if it arises, is small, with its size being 
proportional to the size of quantum fluctuations.

The cubic symmetry of the model at $E_z=0$ is satisfied by the 
quantum-corrected ground-state energy (Fig. \ref{fig:theta}), as the energy  
is invariant under rotations by any angle $\theta=n\pi/6$, where $n$ is any 
natural number. This can be expected as such a rotation corresponds to merely 
directing the orbitals along different cubic axes and/or interchanging
the sublattices $A$ and $B$.

\begin{figure}
      \epsfysize=60mm
      \centerline{\epsffile{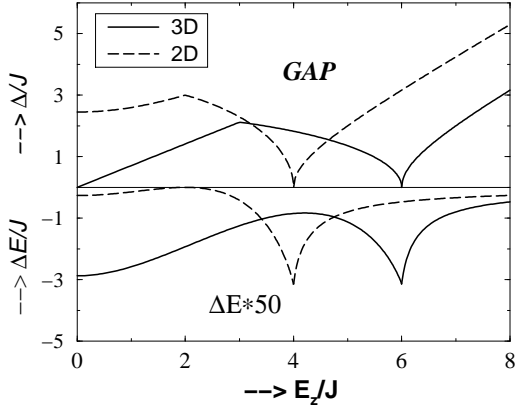}}
\caption
{Gap $\Delta/J$ in the orbital excitation spectrum and 
 energy quantum correction $\Delta E/J$ as
 functions of the crystal-field splitting $E_z/J$.  
 The results for the 3D model and 2D model are shown by full and dashed lines,
 respectively. }
\label{fig:gap}
\end{figure}

\begin{figure}
      \epsfysize=60mm
      \centerline{\epsffile{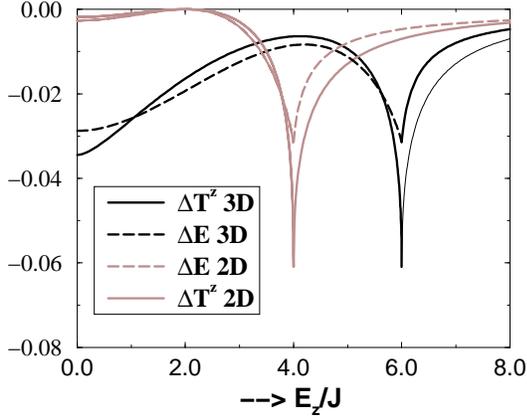}}
\caption
{Quantum corrections to the renormalized order parameter $\Delta T^z$, and
 to the energy $\Delta E$ (in the units of $J$), as functions of the
 crystal-field splitting $E_z/J$. }
\label{fig:ez}
\end{figure}

At finite values of $E_z$ the spectrum becomes gapped (Fig. \ref{fig:gap}) 
because the degeneracy of the classical ground state is removed. Note that 
this is different from a HAF in a finite magnetic field, because there one 
still has full continuous symmetry in the spin-flop phase with respect to 
rotations along the field axis, which justifies the existence of a Goldstone 
mode. For small orbital fields the gap firstincreases up to a maximum reached 
at $E_z=3J$ ($2J$), while for larger values the gap decreases and vanishes at 
$E_z=6J$ ($4J$) for a 3D (2D) model. This can be understood by making the 
analogy with an AF Ising model in a magnetic field. As the field increases, 
the energy of a spin-flip excitation decreases as the loss of interaction 
energy associated with this excitation is partly compensated by the energy 
gain due to the parallel alignment of the excited spin to the field. 
Approaching the phase transition from the spin-flop to FM ordering, this mode 
becomes softer and is found atzero energy exactly at the transition point. 
The increase of the gap for $E_z>6J$ is therefore due to a complete FM 
alignment of the orbitals along the field axis in this parameter regime, and 
any excitation acts then against the orbital field. In Fig. \ref{fig:ez} also 
the quantum corrections to the ground-state energy and the order parameter 
are shown as functions of the orbital field. As expected, the presence of a 
gap in the excitation spectrum suppresses quantum fluctuations.

The quasi-2D state, i.e., the alternation of $|x\rangle$ and $|z\rangle$ on 
two sublattices, is destabilized by any finite orbital field $E_z\neq 0$. 
Taking for instance $E_z>0$, the orbital field shifts downwards the energy 
of the occupied $|x\rangle$ orbitals on one sublattice by $E_z/2$, while the
energy of $|z\rangle$ orbitals increases by $E_z/2$ on the other sublattice,
so that the orbital field does not lower the energy of the system. Therefore, 
the state with $\theta=\pi/4$ is selected instead for the classical ground 
state for any non-zero orbital field\cite{noteez} because in this state the 
pseudospins can tilt towards the orbital field, reducing the energy of the 
system. Such orbitals forming the ground state are shown in Fig. 
\ref{fig:orbitals}. 

\begin{figure}
      \epsfysize=60mm
      \centerline{\epsffile{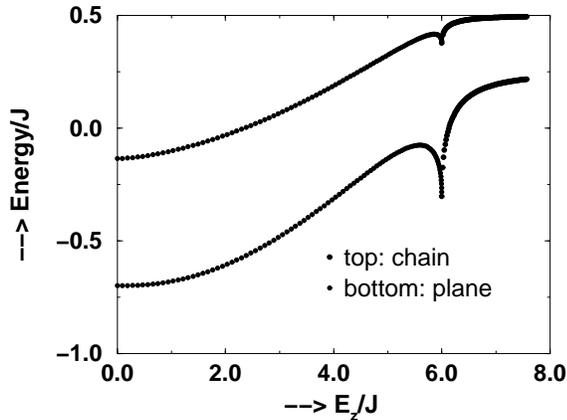}}
\caption
{Energy contribution normalized per one site due the bonds along the $c$-axis
 (chain) and within the $(a,b)$-plane (plane).}
\label{fig:correlation}
\end{figure}

The dispersion for the unrotated state suggests that the effective 
dimensionality of the system is reduced from three to effectively two 
dimensions in the absence of an orbital splitting. This anisotropy results 
in overall smaller quantum fluctuations in a 3D model at $E_z=0$ than in the 
corresponding HAF, as discussed above. The anisotropy of the 3D model 
manifests itself in the energy contribution coming from the bonds in the 
$(a,b)$ plane and along the $c$-axis, as shown in Fig. \ref{fig:correlation}.
For $E_z=0$ the classical energy along the chain is zero, and the orbital
superexchange energy of $3J/4$ is gained by the planar bonds. One finds that 
quantum contributions to the ground-state energy tend to decrease the energy
stored in the bonds along the chain, and increase the energy in the plane.

For the 2D system the situation at orbital degeneracy is quite different (see 
Fig. \ref{fig:modes}). The lack of interactions along the $c$-axis breaks the 
symmetry of the model already at $E_z=0$, opens a gap in the excitation
spectrum and suppresses quantum fluctuations. Increasing the orbital field, 
however, the system resembles the behavior of a 3D system, where also the gap 
closes at an orbital field which compensates the energy loss due to the 
orbital superexchange between identical (FM) orbitals. At this value of the 
field ($E_z=4J$), the full dispersion of the orbital waves is recovered, see 
Fig. \ref{fig:modes}, the spectrum is gapless, and the quantum fluctuations 
reach a maximal value. We would like to emphasize that this behavior is
qualitatively different from the HAF, where the anomalous terms 
$\propto T_i^+T_j^+$ and $T_i^-T_j^-$ are absent, and quantum fluctuations
vanish at the crossover from the spin-flop to FM phase.

\section{Comparison with exact diagonalization in a 2D cluster}
\label{sec:exact}

It is instructive to make a comparison between the analytic approximations of
Sec. \ref{sec:lsw} and the exact diagonalization of 2D finite clusters using 
the finite temperature diagonalization method.\cite{Jak94} In contrast to  
the mean-field approach presented in Sec. \ref{sec:model}, one finds a unique
ground state at $E_z=0$ by exact diagonalization, and no rotation of the 
basis has to be made to investigate the stability of the ground state.
Making the rotation of the basis (\ref{newstatei}) is however still useful 
in exact diagonalization as it gives more physical insight into the obtained
correlation functions which become simpler and more transparent when 
calculated within an optimized basis. Moreover, they offer a simple tool to 
compare the results obtained by exact diagonalization with those of the 
analytic approach, presented in Sec. \ref{sec:numerical}. We shall present
below the results obtained with $4\times 4$ clusters; similar results were
also found for 10-site clusters. 

First we calculate nearest-neighbor correlation function in the ground state 
$\langle \tilde{T}^z_i\tilde{T}^z_{i+R}\rangle$, 
where the operators with a tilde refer to a rotated basis,
\begin{eqnarray}
\tilde{T}^z_i     &=& \cos 2\phi T^z_i     + \sin 2\phi T^x_i \nonumber \\
\tilde{T}^z_{i+R} &=& \cos 2\psi T^z_{i+R} + \sin 2\psi T^x_{i+R},
\label{ttilde}
\end{eqnarray}
so that the correlation function depends on two angles: $\phi$ and $\psi$. 

\begin{figure}
      \epsfysize=80mm
      \centerline{\epsffile{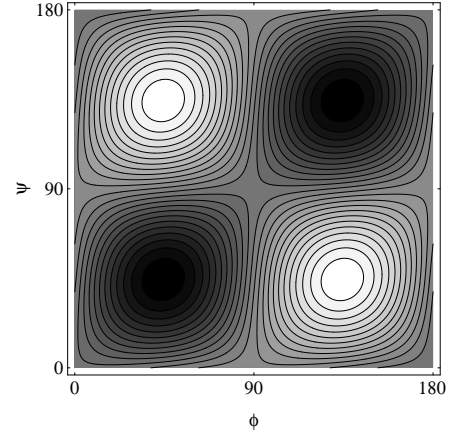}}
\caption
{ Contourplot of the {\em rotated\/} nearest-neighbor orbital correlation 
 function $\langle {\tilde T}^z_i {\tilde T}^z_{i+R}\rangle$ as function of 
 the angles $\phi$ and $\psi$ for a 16-site planar cluster with $E_z=0$ and
 $T=0.1J$. White regions correspond to positive (FM) and black areas to 
 negative (AF) orbital correlations, i.e., 
 $\langle {\tilde T}^z_i {\tilde T}^z_{i+R}\rangle>0.24$ ($<-0.24$), 
 respectively. They are separated by 25 contour lines chosen with the 
 step of 0.02 in the intervall [-0.24,0.24].}
\label{fig:contour}
\end{figure}

In Fig. \ref{fig:contour} the intersite orbital correlation in the ground 
state is shown as a contour-plot. White (dark) regions correspond to 
positive (negative) values, respectively. One finds that the neighbor 
correlations have their largest value if the orbitals are rotated by 
$\phi=\pi/4$ and $\psi=3\pi/4$ (or $\phi=3\pi/4$ and $\psi=\pi/4$), i.e., 
under this rotation of the basis states 
the system looks like a ferromagnet, indicating that the occupied
$(|x\rangle+|z\rangle)/\sqrt{2}$ and $(|x\rangle-|z\rangle)/\sqrt{2}$ 
orbitals are alternating in a 2D model, as shown in Fig. \ref{fig:orbitals}. 
Note that quantum fluctuations are small as in the ground state and one finds
$\langle\tilde{T}^z_i\tilde{T}^z_{i+R}\rangle\simeq 0.246$. This 
demonstrates at the same time the advantage of basis rotation in the exact 
diagonalization study, because in the original unrotated basis one finds 
instead $\langle T^z_i T^z_{i+R}\rangle\simeq 0$, which might lead in 
a naive interpretation to a large overestimation of quantum fluctuations.  

The ground-state correlations $\langle\tilde{T}^z_i\tilde{T}^z_{i+R}\rangle$
are in excellent agreement with the LSWT results. By comparing the results
obtained at $T=0.1J$, $0.2J$ and $0.5J$ we found that the calculated  
low-temperature correlation functions are almost identical in this 
temperature range, and thus the values shown in Fig. \ref{fig:contour} for 
$T=0.1J$ are representative for the ground state. They demonstrate an 
instability of the system towards the symmetry-broken state. This finding is 
similar to the HAF, where such a tendency could also be found, in spite of 
the relatively small size of the considered clusters.\cite{vdL91} 
The intersite correlations decrease at higher temperatures, and thus the 
results at higher temperatures $T>2J$ differ dramatically from those 
presented in Fig. \ref{fig:contour}. By investigating the temperature 
dependence of the orbital correlation functions it has been recently 
established that the orbital order melts at $T\sim J$.\cite{Hor98}

Next we discuss the results for the dynamical orbital response functions
in the case of orbital degeneracy.
The dynamic structure factors for the orbital excitations are defined as,
\begin{eqnarray}
T^{zz}_{\bf q}(\omega)&=&\frac{1}{2\pi}\int^{\infty}_{-\infty} dt\;
\langle T^z_{\bf q} T^z_{\bf -q}(t)\rangle\; \exp(-i\omega t),\\
T^{+-}_{\bf q}(\omega)&=&\frac{1}{2\pi}\int^{\infty}_{-\infty} dt\;
\langle T^+_{\bf q} T^-_{\bf -q}(t)\rangle\; \exp(-i\omega t).
\label{tomega}
\end{eqnarray}
The corresponding orbital response functions evaluated with respect to the
{\em rotated\/} local quantization axis, ${\tilde T}^{zz}_{\bf q}(\omega)$ 
and ${\tilde T}^{+-}_{\bf q}(\omega)$, are defined instead by tilded operators
(\ref{ttilde}). The LSW approximation does not allow to investigate the 
consequences of the coupling of single excitonic excitations to the order 
parameter, represented by the terms $\propto T^x_iT^z_j$, as these terms 
contribute only in cubic order when the expansion to the bosonic operators is 
made. Therefore, the mixed terms involving products of $T_i^z$ and $T_j^x$ at 
two neighboring sites in the Hamiltonian Eq. (\ref{hamorb}) could contribute 
only in higher order spin-wave theory. This motivated us to perform similar 
calculations within finite temperature diagonalization for the simplified 
Hamiltonian without these terms, which we refer to as the truncated 
Hamiltonian.

\begin{figure}
      \epsfxsize=85mm
      \epsffile{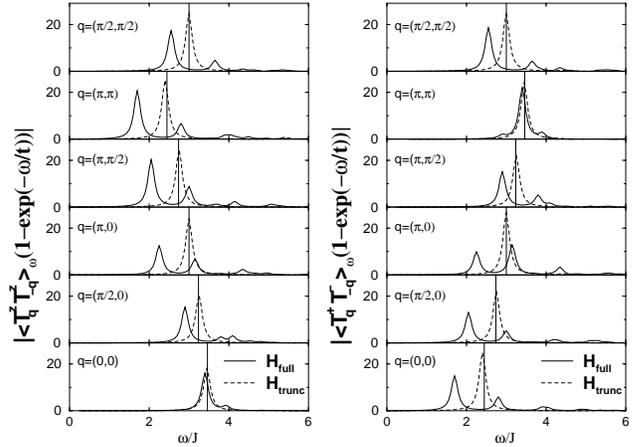}
\caption
{Transverse response function $T^{+-}_{\bf q}(\omega)$ (left part) and
 longitudinal response function $T^{zz}_{\bf q}(\omega)$ (right part) 
 calculated with respect to the original basis states ($\theta=0$) using
 finite temperature diagonalization at $T=0.1J$ for a 16-site planar 
 cluster with $E_z=0$ for:
 (a) the orbital interactions as given by Eq. (\ref{hrotpara}) (full lines); 
 (b) the truncated Hamiltonian (dashed lines). 
 Vertical lines indicate the position of orbital-wave excitations as
 obtained from LSWT. The spectra are broadened by $\Gamma=0.1J$.}
\label{fig:trans}
\end{figure}

The transverse and longitudinal orbital response functions,
$T^{+-}(\omega)$ and $T^{zz}(\omega)$, calculated for the full orbital 
Hamiltonian given in Eq. (\ref{hamorb}), and with respect to the original 
choice of orbital basis $\{|z\rangle,|x\rangle\}$ are shown in Fig.
\ref{fig:trans}. The response functions consist in each case of a dominant 
pole accompanied by a pronounced satellite structure, both for the transverse 
and longitudinal excitations. The dominant pole energies are close to the two 
modes found by the spin-wave analysis, but the energies are lowered for those 
momenta which show strong satellite structures at higher energies. In order 
to establish that these satellite structures are, in spin-wave language, due 
to higher-order processes in spin-wave theory, we performed the same 
calculation for the truncated Hamiltonian. One finds then an excellent 
agreement (Fig. \ref{fig:trans}) between the single modes which result from 
the numerical diagonalization and the position of the orbital-wave as 
determined by Eq. (\ref{eq:omega_phi}).

\begin{figure}
      \epsfysize=80mm
      \centerline{\epsffile{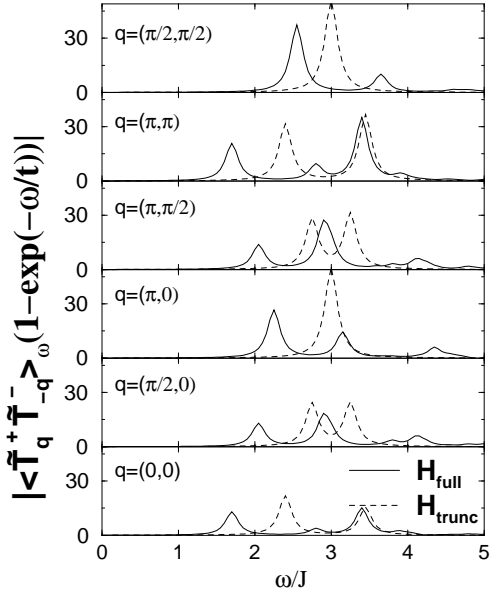}}
\vskip 1cm
\caption
{Transverse response function ${\tilde T}^{+-}_{\bf q}(\omega)$ for the 
 rotated orbitals as a function of frequency $\omega$ for different momenta 
 at low temperature. Calculations were performed for a 16-site 2D cluster at 
 $E_z=0$ and $T=0.1J$ for: 
 (a) the orbital interaction as specified in Eq. (\ref{hrotpara}) 
 (full lines), and 
 (b) neglecting the mixed terms $\propto T^x_iT^z_j$.  
 The spectra are broadened by $\Gamma=0.1J$. }
\label{fig:transrot}
\end{figure}

An additional numerical check was made using the rotated orbital basis 
which determines the transverse correlations 
$\langle\tilde{T}^+_{\bf q}\tilde{T}^-_{-{\bf q}}\rangle$ shown in Fig. 
\ref{fig:transrot}. Here $\phi=\pi/4$ and $\psi=3\pi/4$ for $A$ and $B$ 
sublattice, respectively. In contrast to the unrotated orbital basis, we find 
that the dispersive poles in the longitudinal excitation spectrum vanish 
entirely. Instead, this missing mode appears now in the transverse channel, 
where both modes are found and are accompanied by the satellite structures. 
We note that the response functions in the rotated basis resemble the 
response functions of a HAF, where the spin-wave excitations are also found
only in the transverse response function.\cite{vdL91} As before, the 
satellite structures disappear and two distinct peaks with the same 
intensities are found, if the truncated Hamiltonian is used. These two peaks
merge into a single structure at those values of ${\bf q}$ at which
$\omega_{\bf k}^+(\phi=0)=\omega_{\bf k}^-(\phi=0)$.

\begin{figure}
      \epsfxsize=60mm
      \centerline{\epsffile{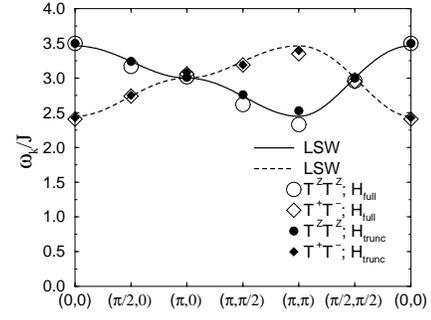}}
\caption
{Dispersion of orbital waves along the main directions in the 2D Brillouin
 zone calculated for a $4\times 4$ cluster. Results for the first moment 
 calculated using the full and truncated Hamiltonian (empty and full symbols) 
 are compared with the dispersions for the two modes as obtained by the LSWT, 
 shown by a solid and dashed line, respectively.}
\label{fig:exact+lsw}
\end{figure}

The successful comparison of the LSWT results with the numerical 
diagonalization is summarized in Fig. \ref{fig:exact+lsw}, where the first 
moment of the spectra of the full Hamiltonian, the single mode which results
from the truncated Hamiltonian, and the dispersion from the LSWT [as found 
from Eq. (\ref{eq:omega_phi}) at $\phi=0$ and neglecting the dispersion due 
to $k_z$], 
\begin{equation}
\omega^{\pm}_{\bf q}=3 J\sqrt{1\pm \case{1}{3}\gamma_{\|}({\bf q})},
\label{wvsq}
\end{equation}
are compared. Clearly, Fig. \ref{fig:exact+lsw} shows an excellent agreement 
between the dispersions calculated within different approaches. From the fact 
that the first moment of the response functions for the full Hamiltonian 
coincides with the response functions of the truncated Hamiltonian, we 
conclude the LSWT approach captures the leading terms in orbital dynamics,
and the satellite structure for the full Hamiltonian is due to higher-order 
orbital-wave interactions that are not taken into account within LSWT.

\begin{figure}
      \epsfxsize=75mm
      \centerline{\epsffile{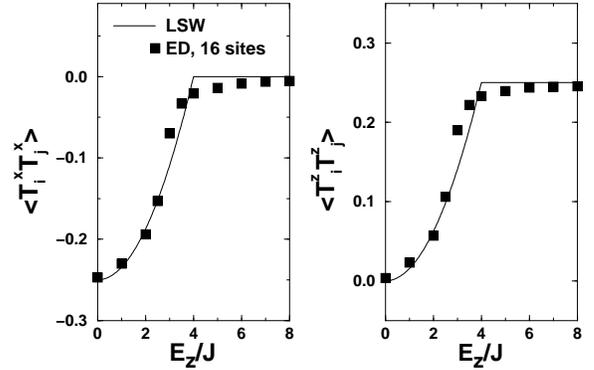}}
\caption
{Ground state expectation values $\langle T^x_i T^x_j\rangle$ (left) and  
 $\langle T^z_iT^z_j\rangle$ (right) for a 16-site 2D cluster ($E_z=0$, 
 $T=0.1J$) as functions of the orbital splitting $E_z/J$ (squares). 
 Full lines 
 represent the same expectation values obtained using the LSWT approach.}
\label{fig:rotation}
\end{figure}

As already discussed in Secs. \ref{sec:model} and \ref{sec:numerical},
the orbitals are tilded in the direction of the effective orbital field due 
to the orbital splitting. This behavior is also found in our diagonalization
studies, and one finds that the expectation values 
$\langle T^z_iT^z_j\rangle$ and  $\langle T^x_iT^x_j\rangle$ depend on 
the orbital splitting $E_z$ (Fig. \ref{fig:rotation}). Both correlation 
functions are calculated using the unrotated basis-set. Although the
crossover to the uniform state with $|x\rangle$ orbitals occupied is smooth
in the numerical study, the agreement with the result of the LWST approach 
is once again very good.

\section{Summary and conclusions}
\label{sec:summa}

In summary we considered the low-energy effective model for strongly 
correlated electrons in two-fold degenerate $e_g$ bands and investigated the 
consequences of the orbital degrees of freedom for a FM system at 
half-filling. Although a saturated ferromagnet might only be realized in
undoped systems in high magnetic fields, we believe that this model is
a relevant starting point in order to understand the influence of orbital 
dynamics in FM doped transition metal oxides, such as the colossal 
magnetoresistance manganites. The FM state can be considered as the reference 
state into which holes are doped.

We have shown that the orbitals order and alternate between the two 
sublattices, which is equivalent to the AF order in the (orbital) pseudospin 
space. The orbital excitation spectrum consists of two branches, one of which 
is found to be gapless in the 3D system within LSWT. This is a consequence of 
the degeneracy of the model at a classical level. Thus, we find here a quite 
peculiar situation. Although the Hamiltonian is not invariant under a global 
orbital rotation $\theta$ [see Eq. (7)], and the correlation functions along 
the different directions change with $\theta$, this rotation does not affect 
the energy of the {\em classical\/} ground state. It turns out, however, that 
quantum fluctuation corrections to the ground state energy {\em restore the 
cubic symmetry\/} and select the rotation angle at either $\theta=0$, or 
$\pi/6$, or $\pi/3$. For these states the orbital excitation spectra are 
purely 2D, which demonstrates that in fact the {\em effective dimensionality 
of the orbital model (\ref{hamorb}) is reduced to two.\/} We note that for a 
gapless 2D dispersion of the orbital excitations in three dimensions one 
expects that orbital long-range order disappears away from $T=0$, like in the 
2D Heisenberg model.

The orbital model (\ref{somfff}) has novel and interesting quantum
properties, and is quite different from the Heisenberg model. In contrast to 
the HAF, the quantum fluctuations in zero field are {\it smaller\/} in the 2D 
than in the 3D orbital model, and the 2D model is therefore {\it more 
classical\/} due to the anisotropy of the orbital interactions. However, a 
finite splitting of the orbitals $E_z$ opens a gap in the excitation spectrum 
of a 3D model, and the quantum effects in an orbital-flop phase stable at 
small values of $|E_z|$ are then reduced. Both in a 3D and 2D case, the 
orbital excitation spectra become gapless at the crossover value of $E_z$ at 
which the 'FM' orbital order sets in. The quantum fluctuations are then 
identical and reach their maximum.

We have verified that the linear approximation within the spin-wave theory
reproduces the essential features of the orbital excitation spectra in the 
present situation, where the total pseudospin quantum number and the $z$th 
component of total pseudospin are not conserved. First of all, the results 
of the LSWT and the first moments of the orbital excitation spectra obtained 
from the finite temperature diagonalization yield two modes in excellent 
quantitative agreement in a 2D model, when the anomalous interactions 
$\sim T^x_i T^z_j$ are neglected. Second, the inclusion of these terms in 
exact diagonalization leads to a satellite structure in the response 
functions, while the first moment remains almost unchanged. In order to 
calculate the satellites observed in the exact diagonalization one needs to 
go beyond the leading order in the spin-wave theory, and include the cubic 
terms in the boson operators. The comparison with finite temperature 
diagonalization allows us to conclude that orbital waves with the dispersion 
of the order of $4J$ are characteristic for the FM states of the undoped 
degenerate $e_g$ systems.\cite{notedisp}

We believe that the presented analytic treatment of orbital excitations 
provides a good starting point for considering the dynamics of a single 
hole in the orbital model. This would allow to understand the origin 
of puzzling optical and transport properties of the doped manganites and 
clarify the expected differences with the hole dynamics in the $t-J$ model.

\acknowledgments

It is our pleasure to thank L. F. Feiner and G. Khaliullin for valuable
discussions, and H. Barentzen for the critical reading of the manuscript. 
One of us (JvdB) acknowledges with appreciation the support by the 
Alexander von Humboldt-Stiftung, Germany. AMO acknowledges partial support 
by the Committee of Scientific Research (KBN) of Poland, 
Project No.~2~P03B~175~14.


\eject
\widetext
\begin{table}
\hspace{0.5cm}
\caption{
Occupied orbitals at two sublattices $A$ and $B$ as found in the classical
FM ground state at orbital degeneracy ($E_z=0$) for a few representative
values of the orbital rotation angle $\theta$ 
[see Eq. (\protect{\ref{newstatei}})]. }
\begin{tabular}{ccc}
   $\theta$   & $i\in A$ & $j\in B$ \\
\hline
      0    & $\case{1}{\sqrt{3}}(3z^2-r^2)\equiv |z\rangle$ &  
             $x^2-y^2\equiv |x\rangle$  \\
  $\case{\pi}{6}$  & $z^2-y^2$   &     $\case{1}{\sqrt{3}}(3x^2-r^2)$   \\
  $\case{\pi}{4}$  &
  $\case{1}{\sqrt{6}}\left[2z^2+(\sqrt{3}-1)x^2-(\sqrt{3}+1)y^2\right]$ &
  $\case{1}{\sqrt{6}}\left[2z^2-(\sqrt{3}+1)x^2+(\sqrt{3}-1)y^2\right]$ \\
  $\case{\pi}{3}$  &  -$\case{1}{\sqrt{3}}(3y^2-r^2)$ &  $x^2-z^2$      \\
  $\case{\pi}{2}$  &  $x^2-y^2\equiv |x\rangle$ &
                      -$\case{1}{\sqrt{3}}(3z^2-r^2)\equiv -|z\rangle$  \\
 $\case{3\pi}{4}$  & 
  $\case{1}{\sqrt{6}}\left[2z^2-(\sqrt{3}+1)x^2+(\sqrt{3}-1)y^2\right]$ &
  $\case{1}{\sqrt{6}}\left[2z^2+(\sqrt{3}-1)x^2-(\sqrt{3}+1)y^2\right]$ \\
\end{tabular}
\label{tabtheta}
\end{table}

\eject
\narrowtext

\end{multicols} 


\begin{references}
\bibitem{notephase}
                The present phase convention is chosen for the electronic 
                Hamiltonian which describes the case of the manganites; the 
                hole case with the opposite phases is described in Refs.
                \protect\onlinecite{Zaa93} and \protect\onlinecite{Fei97}.
\bibitem{noteez} 
                We emphasize that the energies of the states $\theta=0$ and 
                $\theta=\pi/4$ are different due to quantum fluctuations.
                This implies a finite but small orbital field $E_z$ for this 
                reorientation of the ground state, and makes a qualitative 
                difference to the isotropic HAF.
\bibitem{notedisp}
                Note that $J=t^2/U$; instead the dispersion in the HAF for 
                an isotropic hopping $t$ in all three directions is larger
                and amounts to $2zJ$ in the present units, with $z$ being 
                the number of nearest neighbors.
\end{references}
\end{document}